\documentclass[%
 reprint,
 showkeys,
 amsmath,amssymb,
 aps,
]{revtex4-2}

\usepackage{graphicx,subfigure}
\usepackage{bm}

\usepackage[overload]{empheq}

\usepackage{array,tabularx}

\begin{document}

\title{Bifurcation in cellular evolution}

\author{Diego Radillo-Ochoa}
\affiliation{Facultad de Ciencias, Universidad de Colima,
Bernal D\'\i az del Castillo 340, Col. Villas San Sebasti\'an, 
C.P. 28045, Colima, Colima, M\'exico.}

\author{Andrea Rodr\'iguez-Hern\'andez}
\affiliation{Facultad de Ciencias, Universidad de Colima,
Bernal D\'\i az del Castillo 340, Col. Villas San Sebasti\'an, 
C.P. 28045, Colima, Colima, M\'exico.}

\author{C\'esar A. Terrero-Escalante}
\email{Author names are arranged alphabetically.
	Corresponding author: cterrero@ucol.mx}
\affiliation{Facultad de Ciencias, Universidad de Colima,
Bernal D\'\i az del Castillo 340, Col. Villas San Sebasti\'an, 
C.P. 28045, Colima, Colima, M\'exico.}

\date{\today}

\begin{abstract}
Aspects of cell metabolism 
are modeled by ordinary differential equations 
describing the change of intracellular chemical concentrations. 
There is a correspondence between this dynamical system and a complex network.
As in the classic Erd\H{o}s--R\'enyi model,
the reaction network can evolve
by the iterative addition of edges to the underlying graph.
In the biochemical context,
each added reaction implies a metabolic mutation.
In this work 
it is shown that
modifications to the graph topology by gradually adding mutations
lead here too to the formation of a giant connected component,
i.e., 
to a percolation--like phase transition.
It
triggers an abrupt change in the functionality of the corresponding network.
This percolation is mapped into a bifurcation
in the intracellular dynamics.
It acts as a shortcut in biological evolution,
so that
the most probable metabolic state
for the cell
is suddenly switched 
from cellular stagnation
to exponential growth.
\end{abstract}

\maketitle

\section{Introduction}
\label{sec:Intro}

The cell,
considered to be the basic unit of life, 
can be defined
as a self--sustaining chemical system capable of undergoing Darwinian evolution
\cite{benner10}.
No matter how relatively simple a unicellular organism is
when compared to most forms of biological organization,
its metabolism is still a complex system,
i.e.,
it involves a large number of components
subject to nonlinear interactions.
Therefore, 
it is a non trivial task to study 
this whole system or even specific metabolic processes.
In the last few decades,
mathematical modeling has been increasingly recognised as an important tool 
in this area of research.
Models are started to be considered 
as a valuable tool for optimization of bioprocesses, 
to simulate therapy strategies, 
and to predict outcomes in biomedical engineering
\cite{jolicoeur2014}.
Along with these and several others application--oriented uses,
models could also be expected to help to answer fundamental questions
in Cell Biology,
for instance,
what were the basic features
of the structure of the first protocell 
capable of the initial forms of metabolism,
and how complexity and novel functions evolved 
without new mutations compromising already existing metabolic functions.

Starting with the work of Kauffman in 1969 \cite{kauffman69},
models of different aspects of cell metabolism have been proposed 
relying on the fact of the intracellular reactions
conforming a complex network 
\cite{newman2003,barabasi16}
of chemical interactions.
In some of these models
each vertex in a graph is associated with a different 
(not necessarily simple)
chemical species,
having as attribute the corresponding concentration value.
Then, a pair of vertices is connected by an edge if there is a reaction 
that involves the respective species either as reactant or product.
The direction of the edge points towards the vertex corresponding to the product of the reaction.
When reversible reactions are not taken into consideration
an \emph{oriented graph} is used,
i.e.,
a graph having no symmetric pair of directed edges.
In a catalytic reaction network model,
every edge has
as an attribute
the concentration of the catalyst for the reaction it represents.
To complete a model
there are also prescribed
rules for the concentrations to change
according to the intracellular chemical reactions
and the interaction with the external environment.
Along these lines,
several phenomenological models have been thoroughly studied, 
like the simple boolean gene--interaction networks \cite{kauffman69.2}, 
the condensation--cleavage binary polymer model \cite{filisetti14} 
or the evolutionary process based on an artificial chemistry of catalyzed reactions proposed 
in Ref.\cite{jain98}.
Currently,
it has been also realized
that the geometrical representation just described
is not well suited for certain applications, 
as for instance, 
the important task of inferring metabolic pathways.
In such cases,
besides different dynamical rules,
variations of the model may also consider,
for example,
the conservation of the atomic structure of the metabolites during the enzymatic reactions
\cite{arita04},
or using
hypergraphs 
instead of common graphs 
to relate more than one set of reactants and products
\cite{klamt09}.
The procedure for network evolution can also be modified
to include link--deleting or node--deleting
with various kind of restrictions,
or to consider different fitness criteria
\cite{barabasi16,newman2003,jain98,gao2016}.

In the study reported in the present manuscript, 
for the mathematical description of the internal dynamics of a simple cell
we use the catalytic reaction network model 
introduced by Kaneko and Furusuwa in Ref.\cite{furusawa06}.
In the course of the Darwinian--like evolution of this modeled cell,
mutations are added
in each generation
as new reactions in the chemical network. 
Having the largest volume growth rate
among cells of one generation
is considered
as the sole criterion for evolutionary fitness.
Versions of this simulation toolbox have been successful replicating 
properties of real organisms
like differentiation \cite{furusawa00}, 
pluripotency \cite{furusawa09}, 
power--law chemical abundance \cite{furusawa03, radillo2022} 
the reduction in the dimensionality of the phenotypic space changes due to environmental perturbations \cite{furusawa18},
the fact of ‘outward’ central metabolites corresponding mainly to building--block molecules,
the rise of the small--world structure,
and the existence of a small number of keys metabolites exhibiting the highest degree of connectivity
\cite{radillo2022}.

Before continuing with the outline of the motivation and goal of our work, 
let us note that,
though it is customary among researchers
to use interchangeably terms ``network" and ``graph",
we believe it is useful to consider a complex network 
as a pair $(\mathcal{G},\mathcal{F})$,
with the graph $\mathcal{G}(N,L)$ as its fixed geometrical layout
determined by $N$ \emph{vertices} connected by $L$ \emph{edges},
and the functional structure $\mathcal{F}(A,R)$
given by a set $A$ of attributes for the edges and vertices
of $\mathcal{G}$,
as well as a set $R$ of rules according to which these attributes can change.
To emphasize the distinction,
when referring to a network we use \emph{node} 
instead of vertex,
and \emph{link} instead of edge
\cite{newman2003,barabasi16}.
This way, 
we see the graph
as the static topology of a complex network,
while its functionality 
give us the dynamics
that regulates the flow
of nodal information
through the network links.
This separation is particularly relevant
for the aim of our study
because,
as noted in Ref.\cite{CHRISTENSEN2007},
the intracellular dynamics determined by the network functionality 
may be radically different from what seems to be dictated,
for instance,
by the static topology of protein interaction and transcriptional regulatory maps.

Specific details of the version of the \emph{mutation--selection} process 
used here
can be found in
Ref.\cite{radillo2022},
where
four distinct phases of self--organization were unveiled 
during cell evolution
simulated with our toolbox.
In that work it was found that the ‘time' 
when these phases arise
can be parametrized in terms
of the size $N$
of the reaction network.
Around generation $0.5N$ 
the nutrient species prevail as the central reactants of the chemical reactions.
Later, 
by the $1N$--th generation, 
the cell becomes a 
‘small--world',
i.e.,
despite how big the graph can be,
each one of its vertices can be reached from any other one
along paths of edges linking a rather small number of vertices,
usually between 4 and 6
\cite{barabasi16,newman2003}.
Next,
a highly connected core component emerges near generation $1.5N$,
concurrently with the nutrient carriers becoming the central product of reactions.
And finally, after $2N$ generations,
the cell reaches a statistically steady configuration
where the concentrations of the core chemical species 
are described by Zipf's law,
a hallmark of complex systems with a modular structure.
All of these four phases
arise gradually.
The progressive development of
the small--world structure
during the mutation--selection process
is shown in Fig.5 of Ref.\cite{radillo2022}
as the smooth  decrease 
of the mean and standard deviation of the average minimum distance 
between all pairs of vertices.
Similarly,
the closeness centrality 
of both nutrients (outward) 
and carriers (inward) 
raises steadily throughout the evolutionary process
until surpassing the corresponding centrality of the remaining species,
as it is shown in Fig.6 of Ref.\cite{radillo2022}.
As to the formation of
the core component, 
let us first recall that 
it is defined as the subgraph
comprised by the top $20\%-30\%$ 
of the graph vertices ranked with higher overall connectivity.
The main signature of the core
is a positive correlation 
between its nodes connectivity 
and the corresponding species concentration.
This correlation is developed very slowly
over the course of the simulation, 
becoming clearly distinct at late generations
(see Fig.9 of Ref.\cite{radillo2022}),
when the cell is already in the state
of self--organized criticality.

\begin{figure}[!htbp]
\includegraphics{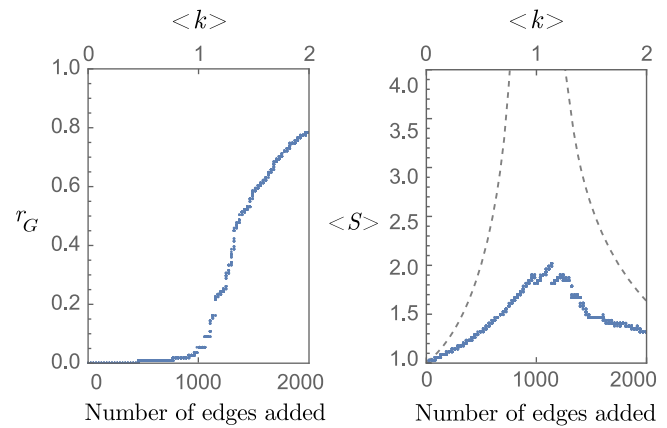}
\centering
\caption[Example of Erd\H{o}s--R\'enyi model realization.]
{Erd\H{o}s--R\'enyi model realization.
Left panel, plot of $r_G$, 
the ratio 
between the number of vertices in the GCC and $N$.
Right panel, 
plot of $\langle S \rangle$,
the average component size
for the finite graph
(blue dots)
and the exact result
(dashed curve).
Quantities are plotted as function of the number of edges added,
and of $\langle k \rangle$,
the average degree of the graph.}
\label{fig:erdos}
\end{figure}

In principle,
the gradual rise of the phases above mentioned
is to be expected, 
as the reaction networks undergoes very little change between generations;
the addition of a single link directly affecting just two nodes.
Nevertheless,
since the pioneer study by Erd\H{o}s and R\'enyi
\cite{erdos60}
it has been shown 
that incrementally adding edges to an initial random graph
leads to the formation 
of 
subgraphs
(called components)
formed by a fraction of the vertices 
that are reachable from each other along a path
in the corresponding graph.
When the largest subgraph with such a structure
comprises almost half of the total number of vertices,
it is called the \emph{giant connected component}
(GCC)
of the graph.
The rise of the GCC 
manifests itself in the form of
a power--law distribution of
the probability of a vertex 
being attached to a given number of edges
(\emph{degree} $k$ of a vertex),
and leads to a continuous phase transition
as shown in Fig.\ref{fig:erdos}.
In the left panel of this figure 
it is plotted the ratio $r_G$
between the number of vertices in the GCC and $N$,
as function of the number of edges added
to an initial set of $N=2000$ vertices,
according to the Erd\H{o}s--R\'enyi model.
As it can be seen,
the first derivative of this plot
seems to undergo a discontinuity
after iteratively adding approximately $1000$ new edges. 
In the right panel
it is plotted the 
average component size
$\langle S \rangle$,
for the same finite graph
(blue dots),
together with the exact result
for this model
(dashed curve)
\cite{newman2003}.
It is now clear
that the discontinuity
occurs
when the average degree of the graph,
$\langle k \rangle$,
equals one,
so that each vertex pair is linked with probability
$1/(N-1)$
\cite{newman2003,barabasi16}.
Both these plots
make manifest, 
in this model of graph growth, 
the familiar to physicists behavior
of continuous phase transitions, 
with the average degree of the graph acting as the control parameter,
the size of the giant component as an order parameter, 
and the average component size behaving like a susceptibility.
Similar transitions have been observed 
in many other models
of graph evolution. 
It can even be made discontinuous
by slightly changing the way the new edges are added
\cite{achlioptas09}.

It is also known that the phase transition
associated with the surge of the GCC
falls in the same universality class 
as percolation
\cite{newman2003,barabasi16,achlioptas09},
the sudden formation of long--range connectivity in random lattices.
Because of the instantaneous increase in the number of links in each chain
of catalytic reactions,
it could imply a drastic change in the reaction network functionality,
perhaps accelerating the emergence of the phases of cellular self--organization
already described in this Introduction,
which in turn seem to be necessary
for the development
of those 
properties of real organisms
which have been theoretically replicated
and were enumerated here--above.
It brings 
to cellular level
the question of whether biological evolution proceeds gradually 
by adding many mutations of small effect 
(Darwinian paradigm)
or in jumps 
due to a few mutations of large effect
(saltational paradigm)
\cite{Theissen2009}.

Taking all of the above into account,
the aim of this work is 
to find out whether there is a percolation--like transition 
in the evolution of the cell as given by the Darwinian--process described 
in Refs.\cite{furusawa06,radillo2022}
and the impact such a phase transition would have in the intracellular dynamics.
This paper is organized as follows.
In the next section
we briefly describe the toolbox used in this research
to simulate cell evolution.
In section \ref{sec:results}
are described the main findings obtained
after having analyzed 
$31$ simulations with networks of three different sizes.
Finally,
in section \ref{sec:concl}
those findings are discussed  
in the light of previously published results,
and conclusions are drawn.

\section{Simulation toolbox}
\label{sec:toolbox}

For simulating cell evolution
a toolbox is often used
which includes
a reaction network model 
and
a Darwinian--like process
\cite{jain98,furusawa03,furusawa06}.
In our study,
the time variation of the concentrations $\mathbf{x}$ of $N$ intracellular chemical species is given by
the following set of ordinary differential equations,
\begin{subequations}
\label{eq:model_ext}
    \begin{align}[left = \empheqlbrace\,]
      \frac{dx_n}{dt} &= - \sum_{j',\hspace{1pt}l'} \sigma_{n, \hspace{2pt} j', \hspace{2pt} l'} \hspace{3pt} x_{n} x_{l'}
+ \hspace{3pt} Dx_{n+4}(\textsc{x}-x_n) \nonumber \\
&- \alpha D x_n \hspace{2pt} 
\sum_{n=0}^3x_{n+4}\left(X-x_n\right)\, ,\\
       \frac{dx_k}{dt} &= \sum_{j,\hspace{1pt}l} \sigma_{j, \hspace{1pt} k, \hspace{0.8pt} l} \hspace{3pt} x_j x_l - \sum_{j',\hspace{1pt}l'} \sigma_{k, \hspace{2pt} j', \hspace{2pt} l'} \hspace{3pt} x_{k} x_{l'} \nonumber \\
&- \alpha D x_k \hspace{2pt} 
\sum_{n=0}^3x_{n+4}\left(X-x_n\right)\, ,
    \end{align}
\end{subequations}
where index $n$ denotes four nutrient species
which show up inside the cell
only by diffusion through the membrane.
Nutrient intake is
induced by a gradient with respect to the environmental concentration $\textsc{x}$,
and tuned 
by the time--dependent product of the diffusion coefficient $D$
and the concentration of the respective nutrient carrier $x_{n+4}$.
Index $k$
stands for the $N-4$ remaining species.
Factors $\sigma_{s,p,c}$ are equal to $1$ 
if reaction $s+c\longrightarrow p+c$ 
takes place,
and $0$
otherwise.
Therefore,
terms in the first line of each equation
describe the 
\emph{net synthesis rate} 
$R_i$.
The second line
accounts for 
 \emph{dilution by growth},
i.e.,
change in concentration $x_i$
due to variation of cell volume
\cite{radillo2022},
\begin{equation}
\label{eq:vol}
\frac{d\ln V}{dt} = \alpha R(t)\, ,
\end{equation}
where $\alpha$ is a proportionality constant
between volume and the total number of molecules,
while
\begin{equation}
\label{eq:sumR}
R(t) \equiv \sum_{i=0}^{N-1} R_i
= D \hspace{2pt} 
\sum_{n=0}^3\left(X-x_n\right)x_{n+4}\, .
\end{equation}
It can be shown that
constraint
\begin{equation}
\label{eq:norm}
\sum_{i=0}^{N-1} x_i(t) = \frac{1}{\alpha},
\end{equation}
must be satisfied along the solutions to Eqs.(\ref{eq:model_ext})
\cite{radillo2022}.

In
Refs.\cite{furusawa06,radillo2022}
are described
all the biologically inspired 
constraints
that determine the possible terms in the right hand sides of Eqs.(\ref{eq:model_ext})
and,
correspondingly,
the topology and functionality of the network
dual to this dynamical system.
Then,
to study the evolution of the dual network
we begin by setting up $m=2N$ cells
sharing an identical internal state $\mathbf{x_0}$,
but
whose graphs are generated by different mutations to a given initial random graph,
with a \emph{mutation}
being the addition of a new edge between two randomly selected vertices.
These $m$ cells are allowed to grow according to Eqs.(\ref{eq:model_ext}),
and
each cell duplication time 
$t_d$
is estimated using relation (\ref{eq:vol})
(see Ref.\cite{radillo2022} for the derivation of 
the specific expression 
and the details on its numerical calculation).
The $n$ cells with the lowest $t_d$ 
are selected to be \emph{mothers} cells of the first \emph{generation}.
Next, 
\emph{daughter} cells are built 
as single mutations
of each one of the mothers.
Every generation is conformed by 
a set of $n \times m$ daughters. 
Subsequently,
daughter cells are allowed to grow, 
and $t_d$ for each one is estimated.
The $n$ ones growing faster up to duplication
are selected as mothers of the next generation,
and so on.

Since
in our
simulation a mutation is the addition of just one link to the network, 
and for each daughter cell it occurs once in each generation,
then 
mutations pile up linearly during the mutation–selection process, 
i.e., 
after evolving up to generation $N$, 
the reaction network of the cells 
has grown in complexity by including $N$ additional catalytic reactions.

\section{Results}
\label{sec:results}

In Ref.\cite{GAMERMANN2019}
was introduced and applied a procedure
to build
the metabolic networks for 3481 real organisms
using data
from the Kyoto Encyclopedia of Genes and Genomes (KEGG)
\cite{kegg}.
Those authors obtained that the number of nodes of the assembled networks
varied between $N=514$ and $N=1154$.
On the other hand,
in Ref.\cite{radillo2022}
it was found that,
throughout the simulation of the evolutionary process,
the reaction networks with $N=250$ nodes
share most of the stages and properties found
for networks with size $N=500$ and $N=1000$.
Taking this into account,
for the present study
were performed a total of 31 simulations:
15 with N=250 nodes,
11 with N=500
and 5 with N=1000,
each one with $20\times N$ cells
per generation.
Taking into account that
by generation $1.5N$
the tightly connected core component in the network 
has usually been already emerged
\cite{radillo2022},
all the simulations were carried out 
up to $1500$ generations.
Even though we parallelized several tasks
to run our simulations in multicore architectures,
together with the analysis of the outcomes,
they still accounted for more than a thousand of computing hours.

Despite the high degree of complexity of the simulations,
we found the outcomes to be generic,
and qualitatively independent of $N$,
as well as of the values of the other
parameters in Eqs.(\ref{eq:model_ext}).
For the findings summarized in table \ref{table:table1}
\begin{table*}
\begin{tabularx}{0.8\textwidth} { 
  >{\setlength\hsize{1\hsize}\raggedright\arraybackslash}X 
  >{\setlength\hsize{.5\hsize}\raggedleft\arraybackslash}X 
  >{\setlength\hsize{.5\hsize}\raggedleft\arraybackslash}X 
  >{\setlength\hsize{.5\hsize}\raggedleft\arraybackslash}X }
 \hline
 Network size ($N$) & \textbf{250} & \textbf{500} & \textbf{1000}\\
 \hline
 Number of simulations performed. & 15 & 11 & 5\\
 Number of cases with transient regime. & 15/15 & 10/11 & 5/5\\
 Generation at which the transition begins ($g_i$). & $81\pm18$ & $155 \pm 50$ & $381 \pm 168$\\
 Generation at which the transition ends ($g_f$). & $139 \pm 79$ & $243 \pm 89$ & $452 \pm 286$\\
 Power--law exponent before and after transition. & -7, -0.85 & -7, -1 & -7, -1\\
 Interval of generations at which nutrients become central substrates (zeroed at $g_f$). & $[-5g,+10g]$ & $[-5g,+5g]$ & $[-4g,+1g]$\\
 Effective control parameter ($\beta_\text{eff}$). & $1.64 \pm 0.18$ & $1.69 \pm 0.24$ & $1.66 \pm 0.33$\\
 \hline
\end{tabularx}
\caption{Results summary.
Thirty--one simulations were performed with networks of three different sizes.
In the third row it is shown that in all but one the rapid growth of the $GCC$ was observed.
For these, 
the transient regime starts at $g_i\approx 0.5N$,
ends at $g_f<1N$,
and the corresponding network was sensitive to an external stimulus.
In the 6th row
are displayed the median
of the exponents of the power--law for the rank--size distribution of the concentrations
before and after the transition,
showing a clear jump in the values.
In the next row it can be observed that
the nutrients usually dominate the outward centrality
a few generations
around $g_f$.
In the last row it is shown the value 
of the effective control parameter
$\beta_\text{eff}(g_f)$.
}
\label{table:table1}
\end{table*}
it was used
$D=6$,
$X=0.2$
and
$\alpha=1$.
In all but one simulation a phase transition was detected.
The results detailed below correspond to these 30 cases.

\subsection{Percolation in the reaction network}
\label{ssec:percolation}

The mutation--selection process 
is started with random graphs 
that had,
on average,
one
edge per vertex.
Because of the sparse nature of the initial graph,
to be able to clearly observe the giant component,
we applied a procedure similar to the one used in
Ref.\cite{CHRISTENSEN2007}
and analyzed an effective network 
derived from the original reaction network
by neglecting all nodes with concentrations below a threshold value.
Since the $x_i$ 
are constrained by Eq.(\ref{eq:norm}) to be less than one
and always enter as quadratic terms in Eqs.(\ref{eq:model_ext}), 
those species 
with concentrations below $10^{-4}$
at the duplication time
usually played a negligible role
throughout the modeled cell lifetime.

\begin{figure*}[!htbp]
\includegraphics{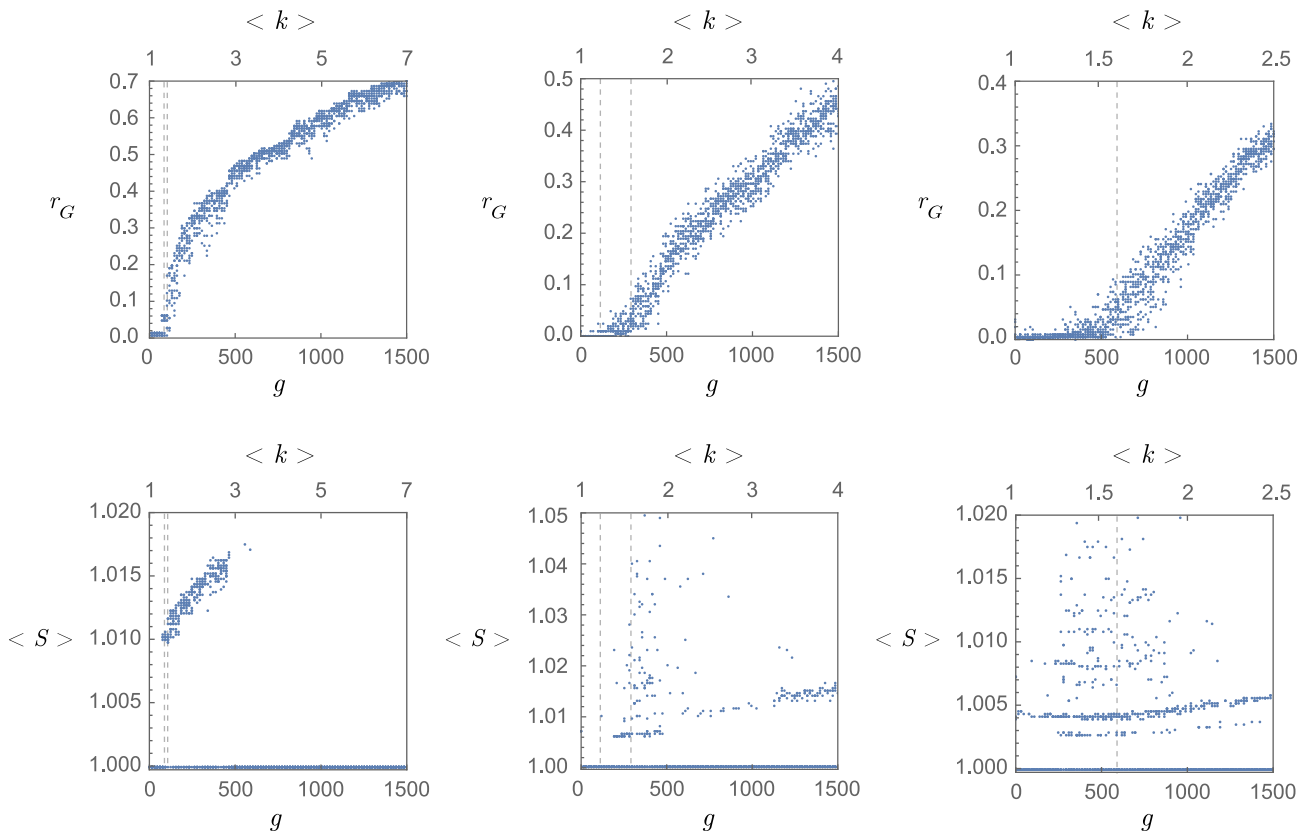}
\centering
\caption[Rise of the giant component.]
{Rise of the giant component.
Top panel, plots of the ratio $r_G$ during three simulations of cell evolution
using reaction networks 
with $N=250$ (left), $N=500$ (center) and $N=1000$ (right) species.
Bottom panels, corresponding plots of $\langle S \rangle$.
In the horizontal axis $g$ stands for the simulation generation.
Vertical dashed lines enclose the transient regime.
}
\label{fig:gscc}
\end{figure*}

In the top panels of Fig.\ref{fig:gscc}
are shown generic examples
of the formation of the giant component
in our simulations.
For the effective graph of the fittest cell
of each generation $g$,
the ratio $r_G$
is plotted.
In the bottom panels is presented the corresponding behavior
for the average component size, 
$\langle S \rangle$.
As it can be seen by comparing figures \ref{fig:erdos} and \ref{fig:gscc},
there are similarities 
and differences between the behavior of our evolutionary process
and that of the classic Erd\H{o}s--R\'enyi model.
The most important similarity is that
in both cases
we are able to note a clear change in the growth trend of the GCC.
As to the obvious differences observed,
it must first be considered that,
since we are dealing with directed graphs,
we have to take the direction of the paths into account
and look instead for the giant \emph{strongly} connected component
\cite{newman2003,barabasi16}.
Secondly,
even one--thousand vertices 
happen to be rather few 
for capturing some features characterizing
graphs in the limit of large $N$,
in particular,
the statistics of the graph components,
excluding the GCC
\cite{newman2003}.
Furthermore, 
in most models
it is
followed the evolution of a single graph,
while we follow the graph of the fittest cell 
amongst the $20\times N$ cells of each generation.
Mother--daughter trails can come to a dead--end
at any point
in our simulation,
and
the fittest cell of one generation
does not necessarily grows faster than the fittest one of the previous generation.
All these differences are reflected as a scattering of points in the plots,
instead of the rather plain curves observed for the Erd\H{o}s--R\'enyi model
in Fig.\ref{fig:erdos}.
Hence, 
the onset of the rapid growth of $r_G$
does not correspond here to a point,
but to a transient regime
along which 
builds up
the amount of cells in each generation
having the topology needed for the rise of the giant component.
In table \ref{table:table1}
it can be observed that,
as in the examples of Fig.\ref{fig:gscc},
this regime typically starts by generation $0.5N$ 
and ends well before generation $1N$.
It happens when 
the average number of edges per vertex in the effective graph
is above $1$
(recall that we start our simulations with $\langle k \rangle =1$),
but still less than $2$.

\begin{figure}[!htbp]
\includegraphics{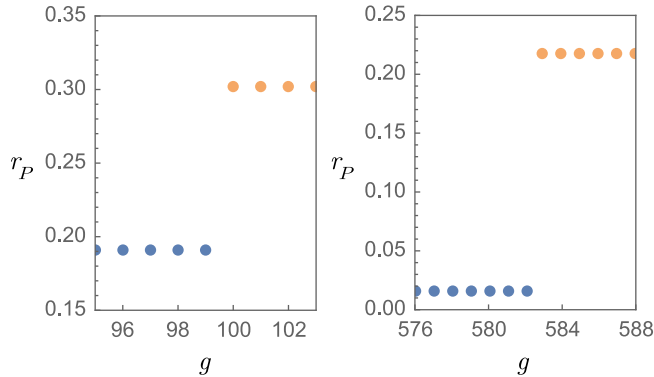}
\centering
\caption[Response to doubling the external nutrients.]
{Fittest cell response to doubling the environmental concentration of nutrients. 
In these two simulations were used reaction networks with $N=250$ (left) and $N=1000$ (right) species. 
The transient regime ends at $g_f=99$ and $g_f=582$, respectively.
On the vertical axis, $r_P$ is the ratio between 
the number of non--inert nodes
and $N$.}
\label{fig:perco}
\end{figure}

Results in Fig.\ref{fig:gscc} 
imply that,
in our case,
the size of the GCC does not act as a distinct order parameter, 
as it does in most models of random graph evolution 
and,
consequently,
the average component size does not longer behave like a susceptibility.
Nevertheless,
we verified that 
it is indeed observed a continuous phase transition.
As it was mentioned earlier,
for many models of graph evolution
the surge of the giant component
is equivalent to a second order transition in a lattice.
In our case,
this percolation--like transition causes
a radical change in the reaction network functionality.
To uncover this, 
we study the network response to external stimuli.  
With that aim
we took the fittest cell of each generation,
doubled the environmental concentration of nutrients $X$,
solved system (\ref{eq:model_ext}) numerically
with the initial conditions
set to $x_i(t_d)$,
and then counted the number of non--inert nodes,
i.e.,
those whose concentrations,
after a time much larger than $t_d$,
varied by an amount greater than $10^{-6}$.
In Fig.\ref{fig:perco}
is plotted
the ratio $r_P$ between 
the number of non--inert nodes
and $N$,
as it varies in an interval of generations
centered at the end of the transient regime.
For the examples presented in this figure
it is clearly observed a discontinuity in the trend of the values of $r_P$
before and after
the transient regime ends.
Again,
it must be emphasized that the data
do not correspond to one and the same graph evolved from an original random graph,
but to the fittest among the $20 \times N$ cells of each generation.
As summarized in the sixth row of table \ref{table:table1},
a similar effect of doubling $X$
was observed in all of the cases 
where there was a rapid rise of the GCC.

\begin{figure}[!htbp]
\includegraphics{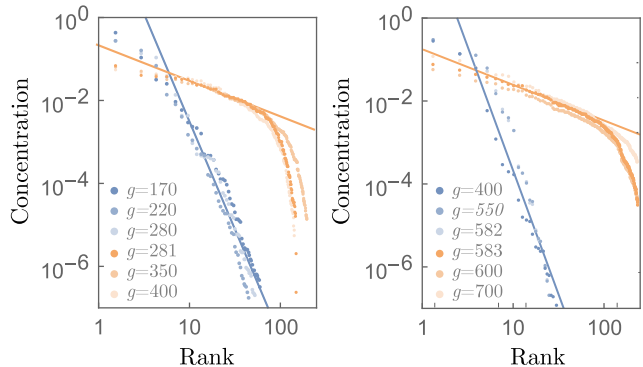}
\centering
\caption[Ranked distributions of concentrations.]
{Rank--size distributions for concentrations of the fittest cell. 
In this simulation were used reaction networks with $N=500$ (left) and $N=1000$ (right) species.
The transient regime ends at $g_f=280$ and $g_f=582$, respectively.
The slope of the blue line is $-7$,
and for the orange line, $-1$.}
\label{fig:pl250}
\end{figure}

More evidence on the nature of this transition
is presented in Fig.\ref{fig:pl250}
where it is shown,
in other examples,
that it also causes a sudden jump in the value
of the exponent of the power--law
describing the rank--size distribution for the asymptotic concentrations
of the fittest cell.
This discontinuity
implies an abrupt change in the statistical properties of $x_i$.
For all these generations
the distributions
can be fitted by power--laws
\footnote{The exponents are found using the Python library plfit.py, 
an implementation by Adam Ginsburg of the general algorithm presented in Ref.\cite{clauset09}.}
but,
as it can be seen in the seventh row of table \ref{table:table1},
exponents 
before the critical point
(approximated by the blue line slope)
correspond to data with finite variance,
while this is not true for the distributions after the transition
(orange line).

\subsection{Bifurcation in intracellular dynamics}
\label{ssec:bif}

So far we have analyzed the impact of the
mutation--selection process
on the topology and functionality of the network describing the fittest cell.
But to the geometrical representation of a reaction network
corresponds a specific dynamical system given by Eqs.(\ref{eq:model_ext}).
A meaningful insight 
into the cellular metabolism is revealed
now
if this 
\emph{dynamical system--complex network} 
duality 
is used 
to uncover 
what the percolation--like transition
implies for the dynamics
of the intracellular reactions.

Let us start by recalling
that a dynamic steady state of the cell metabolism is reached when,
for each chemical species,
the net synthesis rate $R_i$ is balanced by dilution by growth,
so that,
\begin{equation}
\label{eq:eq}
\frac{d{x}_i}{dt} = 0\, , \quad \forall i\in \{0,1,\cdots,N-1\}.
\end{equation}
The following proposition comes in handy:
\emph{A cell is at a dynamic steady state
if,
at any time,
the finite ratio 
between the concentrations of any two intracellular species
is equal 
to the ratio 
between the corresponding net synthesis rates. 
This implies that at this state 
all the species are diluted steadily}.
To prove this statement,
let us start by noting
that the origin 
$\{x_i(t)=0\, , \forall i$\},
is not an equilibrium of this dynamics,
because is not consistent with constraint (\ref{eq:norm})
for any finite $\alpha$.
Then,
according to our proposition,
\begin{equation}
\label{eq:sstate1}
\frac{x_i(t)}{x_j(t)} = \frac{R_i(t)}{R_j(t)}\, , \quad \forall (i,j)\, ,
\end{equation}
and $x_j\neq 0$.
Let us now sum over index $i$ and use constraint (\ref{eq:norm})
and expression (\ref{eq:sumR})
to obtain
\begin{equation}
\label{eq:sstate2}
\frac{1}{\alpha x_j(t)} = \frac{R(t)}{R_j(t)}\, , \quad \forall j\, ,
\end{equation}
which is consistent with Eqs.(\ref{eq:model_ext})
only if conditions (\ref{eq:eq}) are fulfilled.
Finally,
taking the derivative of Eq.(\ref{eq:sumR})
it can be verified that if conditions (\ref{eq:eq}) are met,
then the volume change rate remains constant.
It can be positive (exponential growth), 
zero (stagnation) 
or negative (atrophy).

\begin{figure*}[!htbp]
\includegraphics{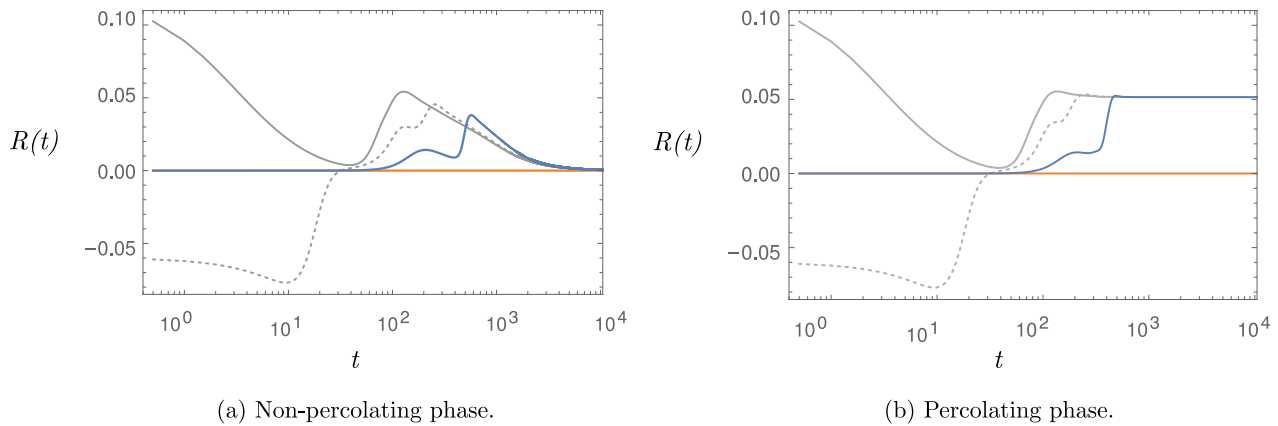}
\centering
\caption[Volume rate of change.]
{Volume rate of change, $R(t)$, as function of time
for generations $98$ (a) 
and $99$ (b)
of the simulation in the left panel of Fig.\ref{fig:gscc}.
The orange curve corresponds to a stagnation steady--state, $R(t)=0$.
The solution displayed with a blue curve starts as a tiny perturbation of the stagnation state.
The continuous and dashed gray curves correspond to solutions starting with $R(0)>0$ and $R(0)<0$, respectively.} 
\label{fig:dyn250}
\end{figure*}

Now,
since in our study we consider several hundred species,
due to the huge dimension of the corresponding phase--space
and the nonlinearity of Eqs.(\ref{eq:model_ext}),
it is nearly impossible to carry out a qualitative analysis of their solutions,
not even
of dynamic steady states.
Still in lower dimensions
there is an infinite number of ways
of satisfying conditions (\ref{eq:eq}),
as they often define hyperplanes and not
just isolated points in the phase--space.
Moreover,
the classification of the stability
of any of the possible isolated equilibria
is also nontrivial 
because the associated flow cannot be linearized
in their vicinity
\cite{guckenheimer83}.
Therefore,
very little can be inferred about the modeled cellular metabolism
from the information on the matching phase--portrait.
Fortunately,
having analyzed,
for instance,
any of the examples presented in the previous subsection,
we can now solve numerically the ODE systems
corresponding to the networks
immediately before and after that transition,
and study the behavior of the solutions.
For the case with $N=250$ in Fig.\ref{fig:gscc},
several of these solutions are represented in Fig.\ref{fig:dyn250},
where $R(t)$ is plotted
instead of each one of the ${x_i(t)}$.
Whenever we observed $dR(t)/dt=0$,
conditions (\ref{eq:sstate1})
were tested to find out whether
it corresponds to a dynamic steady state.
Recall that
the meaningful for us dynamics
takes places around the origin of the phase--space,
which,
as mentioned before, 
is not an equilibrium itself.
In Fig.\ref{fig:dyn250} (a)
the system
corresponds to the network
immediately before the percolation--like transition.
For this case
it can be observed that 
stagnation,
$R(t)=0$,
is always the attractive asymptotic state.
In
Fig.\ref{fig:dyn250} (b)
is plotted $R(t)$
in the percolating phase.
Stagnation states 
still persist,
but they are now unstable.
Moreover,
a new attractor arose, 
a steady state
where the cell grows exponentially.
To check that this state does not exist
before the transition
we solved system (\ref{eq:model_ext})
backward and forward in the non--percolating phase
with initial conditions set at the 
$R(t)> 0$
attractor
of the percolating phase,
and no convergence to any other state, 
but to $R(t)=0$, 
was detected.

Adding a network link
corresponds to considering
new reciprocal terms
$x_p x_c$  and $x_{s} x_{c}$
in the right hand sides
of Eqs.(\ref{eq:model_ext}),
equivalent to switching the values 
of the related coefficients
$\sigma_{s, \hspace{1pt} p, \hspace{0.8pt} c}$
from zero to one.
The exchange in the stagnation stability 
and the emergence of the new stable equilibrium
imply that
the initial phase--portrait
is not longer topologically equivalent
to the one resulting
after
switching on
the couple of parameters
$\sigma_{s, \hspace{1pt} p, \hspace{0.8pt} c}$.
This is the distinct signature of what is known as
\emph{bifurcation} of a vector field
\cite{guckenheimer83}.
Even if 
these parameters are switched on randomly,
for this bifurcation to happen
it should be done so
that it gradually increases the degree of coupling
between Eqs.(\ref{eq:model_ext}).
Obviously,
this can be done in several ways,
i.e.,
there are many paths leading to this bifurcation.
We found that,
independently of the system dimension,
at the critical number of switched on 
$\sigma_{s, \hspace{1pt} p, \hspace{0.8pt} c}$,
if all $x_i$ below a threshold value are neglected,
the probability of any pair of species
being involved in the same reaction is approximately
$1/(N-1)$. 

\section{Discussion}
\label{sec:concl}

Let us start the discussion of our results by 
recalling that we have
found that the number of generations 
at which take place 
the events described in this manuscript
can be parametrized
by $N$, 
the number of species in the modeled cells. 
In particular $N$ takes here the values 250, 500 and 1000,
the last two values been consistent
with what have been found for the metabolic networks 
of real organisms
\cite{GAMERMANN2019},
while for $N=250$
it was found previously
\cite{radillo2022}
that
reaction networks with this number of nodes
share most of the properties found
for larger networks.

We have reported in this manuscript 
evidences 
that including new chemical interactions
during the evolution of the intracellular reaction network
leads to the rapid growth of a giant connected component.
Differently to what happens in other models of graphs evolution,
here this event does not takes pĺace at a specific iteration,
but rather after a transient regime.
Though it
varies from case to case,
this regime is typically centered around generation $0.5N$.
As mentioned before,
this is the same generation when usually
the nutrients clearly become the central reactants of the chemical reactions
\cite{radillo2022}.
As already noted, 
this last event occurs gradually,
while the sudden rise of the giant strongly connected component
is ubiquitous in the evolution of random directed graphs.
Therefore,
it is the functionality 
and not the topology
of the reaction network
what induces this coincidence;
at this stage of the simulation,
the mutations lead to the fast increase 
of the length of the chemical chains
in all cells,
but the evolutionary process selects
as the fittest ones those
where nutrients happen to be at the origin
of most pathways,
because in those cases 
the nutrients intake 
keeps the concentrations of the involved species
above a critical value.
The fact that a similar topology
does not necessarily implies
a unique functionality
was,
for instance,
discussed in Ref.\cite{CHRISTENSEN2007},
where it was noted that
the dynamic of a cellular network 
may be radically different 
from what is suggested 
by the given topology of a network
of protein interaction and transcriptional regulatory maps.
On the other hand,
since it is considered that the largest component
of a graph
becomes its giant connected component
when it comprises about half of the vertices,
it means that,
by the time the giant component starts to form,
its complement remains mostly disconnected.
This way,
as observed in our simulations,
the end of the transition period
should precede the setup,
around generation $1N$,
of the small--world structure observed in 
Ref.\cite{radillo2022}.

We have shown that,
indeed,
the fast growth of the giant strongly connected component
signals a percolation--like phase transition
in the reaction network functionality.
This percolation is characterized by a sudden change 
in the cell response to an external stimulus;
after duplicating the nutrients external concentration,
the percent of species actively participating in the chemical reactions
is usually increased by a factor of 2.
Ironically,
the percolation
also leads to concentration statistics 
described by power--laws with infinite variance,
i.e.,
allowing for a few species to account
for most of the chemical content of the cell.
It is obvious that intracellular chemical concentrations
cannot actually have infinite variance and,
according to constraint
(\ref{eq:norm}),
one would expect
$x_i \propto \mathcal{O}(1/N)$ for all $i$.
This could be the case before the transition
but,
in the percolating phase, 
cell metabolism becomes capable of
‘black swan' behavior
(see Ref.\cite{furusawa03} for an example in real microorganisms).

In the continuous phase transition 
characterizing many other models of graph evolution
the size of the giant component is the order parameter,
nevertheless,
in our simulations
the variation of this quantity
is not monotonic
so, 
there seems not to exist a clear--cut order parameter
for the observed by us phase transition.
This is consistent with the difference
between physical and biological systems
found in the experimental study reported in Ref.\cite{GIULIANI2004}.
Because of the interaction
between its elements,
the growth of complexity
during the evolution
of a biological system
is usually aimed to
benefit most of its components,
and the solution of this problem of optimization with constraints is commonly not unique.
In the case of a cell,
its evolution could have proceed through several lanes,
in such a way that,
at any given state of the mutation--selection process,
the same “metabolic status” of the system is reached following different preferred reaction pathways
\cite{GIULIANI2004}.

In that regard,
using the correspondence 
between 
the geometrical and dynamical representations
of a reaction network,
we found out
that the percolation--like phase transition
is mapped into
a bifurcation of the vector field
describing the intracellular dynamics.
Starting from a simple cell
with a random and sparse structure of catalytic reactions,
a Darwinian process
causes an early qualitative switch
in the cellular metabolism,
so that the fittest cell
suddenly stop being
the one
which, 
just by chance,
happens to be as far as possible from stagnation,
to become the one 
naturally closer to the state of exponential growth.
Relation between
phase transitions and bifurcations
has been documented before
(see for instance chapter 10 in Ref.\cite{Sornette2006}
and references therein).
Nevertheless,
to uncover it
the behaviour of a complex system
is usually approximated  with 
a one--dimensional nonlinear dynamic equation
for the average over time and space of the order parameter,
with the particular property that, 
as the control parameter crosses the critical value for the phase transition,
the corresponding $1D$ dynamical system undergoes a bifurcation.
Since in our case we do not have well defined control and order parameters,
that approach seems to be useless.
A closer affinity with our results
have those reported in Ref.\cite{gao2016},
where a formalism was devised
to reduce to an effective $1D$ ordinary differential equation
the dynamics of multi--dimensional systems consisting of
a large number of components that interact through a complex network.
In particular, 
applying their method
to the transcription networks of 
{\it Saccharomyces cerevisiae} 
and 
{\it Escherichia coli}, 
the authors found that,
in the effective phase--space,
the cell undergoes a transition from death to a resilient state,
though nothing is said about
what kind of phase transition in the functionality 
of the original complex network 
the bifurcation corresponds to.
We estimated the value at the end of the transition regime
of the effective control parameter
$\beta_{\text{eff}}$
proposed by these authors
and reported it in the last row of table
\ref{table:table1}.
Even if the statistics are too poor to make definite conclusions,
$\beta_{\text{eff}}$
seems to be around a fixed value,
regardless of the specific topology
and functionality of the analyzed network,
a universality noted in Ref.\cite{gao2016}.
Besides the fact that we have found a bifurcation 
of the full $N$--dimensional phase--space,
another difference between those and our results
seems to be the cell states before and after the transition.
Before the bifurcation reported by us
the modeled cell 
will experience a decelerated growth,
but it will be eventually able to duplicate.
Of course,
in a more realistic setup,
prolonging the duplication time
makes a cell liable 
of dying before dividing because of,
for instance,
mechanical trauma, extreme thermal effects, ischemia,
toxins and pathogens inflow.
To the contrary,
growth after the bifurcation here described
would be robust to 
metabolic changes
caused by environmental or internal factors
not considered in this model,
provided these changes
occur at time scales larger than
the now relatively short duplication time
characteristic of the exponential growth.

The drastic change in cellular metabolism
induced by a minor modification in its structure
can be seen as a shortcut in evolution.
Besides making the cell fairly resilient to some adverse external influences,
it also enables it to reach faster 
the level of complexity required for
new processes to take place
\cite{filisetti14,jain98,furusawa00,GIULIANI2004},
such as the emergence
of small--world and hierarchical structure,
of auto--catalytic cycles
and 
of a dominant core of metabolites,
which in turn allow for
immature and non--specialized cells
to develop into mature ones
with specialized forms and functions,
and able
to produce several distinct biological responses
individually
or as a 
collective behavior.
If such a bifurcation in the intracellular dynamics
could actually occur,
then,
along with Darwinism and saltationism
\cite{Theissen2009},
there would be another way for biological evolution to proceed,
at least at cellular level:
the gradual accumulation of mutations of negligible effect
leading to a sudden change in the phenotype.
Since there would be many paths leading,
through the bifurcation,
to the same metabolic status,
then the new phenotype would not be exclusive of a
\emph{hopeless monster},
but rather an inevitable outcome of evolution.

We would like to conclude by noting
that,
since in our study we are dealing with universal properties of complex systems,
it makes sense 
to conjecture 
that 
\emph{there is a wide class of phenomena
modeled by complex networks
of size $N$
where 
topology--induced
phase transitions in the network functionality 
are mapped into bifurcations of a dual 
$N$--dimensional vector field}.
If true,
this duality 
will be reinforced as a powerful tool
for studying a class of high--dimensional non--linear dynamical systems,
where usual analytical techniques are fruitless,
if suitable at all.

\section{Acknowledgments}

We thank Vrani Ibarra and Roberto S\'aenz
for useful discussions.
We also acknowledge HypernetLabs and Google Cloud Services 
for allowing us to use their computational facilities to run most of our simulations.
The work of A. R-H and D. R-O was supported by CONACyT grants for graduate studies.

\bibliography{references}

\end{document}